\documentclass{aa}
\usepackage{graphicx}
\usepackage{natbib}
\usepackage{txfonts}
\usepackage{color}

\def\aap{Astron. Astrophys.}

\def\apjl{Astrophys. J. Lett.}

\def\apjs{Astrophys. J. Suppl. Series}

\begin{document}
   \title{Nonlinear force-free modelling: influence of inaccuracies in
   the measured magnetic vector.}

   \subtitle{}

   \author{T. Wiegelmann \inst{1}
          \and
         L. Yelles Chaouche \inst{1}
         \and
          S. K. Solanki \inst{1,2}
          \and
          A. Lagg \inst{1}
\fnmsep\thanks{}
          }

   \offprints{T. Wiegelmann}

   \institute{Max-Planck-Institut f\"ur Sonnensystemforschung,
Max-Planck-Stra\ss{}e 2, 37191 Katlenburg-Lindau, Germany\\
              \email{wiegelmann@mps.mpg.de}
              \and
 School of Space Research, Kyung Hee University, Yongin, Gyeonggi 446-701, Korea}

   \date{}


  \abstract
   {
   Solar magnetic fields are regularly extrapolated into the corona starting
   from photospheric magnetic measurements that can suffer from significant uncertainties.
   }
   {
   Here we study how inaccuracies introduced into the maps of the photospheric magnetic vector
   from the inversion of  ideal and noisy Stokes parameters influence the
   extrapolation of nonlinear force-free magnetic fields.}
  {
  We compute nonlinear force-free magnetic fields based on simulated vector magnetograms,
  which have been produced by the inversion of Stokes profiles, computed from
  a 3-D radiation MHD simulation snapshot. These extrapolations are compared with
  extrapolations starting directly from the field in the MHD simulations, which is
  our reference. We investigate how line formation
  and instrumental effects such as noise, limited spatial resolution and
  the effect of employing a filter instrument influence the resulting magnetic field structure. The
  comparison is done qualitatively by visual inspection of the magnetic field distribution and
  quantitatively by different metrics.
 }
   {The reconstructed field is most accurate if ideal Stokes data are inverted
   and becomes less accurate if instrumental effects and noise are included.
   The results demonstrate that the non-linear force-free field extrapolation
   method tested here is relatively insensitive to the effects of noise in measured
   polarization spectra at levels consistent with present-day instruments.
   }
   {
   Our results show that we can reconstruct the coronal magnetic field as a nonlinear
   force-free field from realistic photospheric measurements with an accuracy of a few
   percent, at least in the absence of sunspots.
   }

   \keywords{Sun: corona --
                Sun: magnetic fields --
                Sun: photosphere}
 \authorrunning{Wiegelmann et al.}
 \titlerunning{Force-free modelling: influence of spectral lines}
   \maketitle
%
\section{Introduction}
Except for a few individual cases, e.g., in newly developed active regions
\citep{solanki:etal03}, we cannot measure the full magnetic vector in the
solar corona at high resolution directly and we have to rely on
extrapolations from photospheric measurements.
\cite{wiegelmann:etal05} compared the direct magnetic field measurements by
\cite{solanki:etal03} with extrapolations from the photosphere. This work
revealed the importance of field aligned electric currents for an accurate
magnetic field reconstruction and the need for photospheric vector
magnetograms as boundary data. The photospheric data suffer from a number of
inadequacies, however, whose influence on the quality of the extrapolations
needs to be studied in detail. Firstly, the magnetic field in the photosphere
is far from being force-free, \citep[see, e.g.,][]{metcalf:etal95}, which
leads to significant errors if used directly for a force-free magnetic field
extrapolation \citep[see][for details]{metcalf:etal08}. These can, however,
be greatly reduced by appropriate preprocessing \citep{wiegelmann:etal06}.
Another well known problem is that the noise level of the magnetic field
transverse to the line of sight is typically more than one order of magnitude
higher than for the longitudinal component deduced from the Stokes parameters of
Zeeman split spectral lines.
An additional complication arises for vector magnetic field measurements
made far away from the disk center, where the vertical and line-of-sight
field are far apart \citep[see][for details]{gary:etal90}. In this case
one has to use also the transverse field  to derive the vertical field.
\cite{venkatakrishnan:etal89} showed that the increased noise by using the
transverse field is tolerable for heliocentric distances of less than $50^\circ$.
For our investigations
we assume that the observations are made close to the disk center, where the vertical
magnetic field is almost identical with the line-of-sight field and the horizontal
component is almost identical with the transverse field. Consequently our results
are not directly applicable to regions observed for heliocentric distances of
larger than about $50^\circ$, where the noise in the transverse field influences
significantly the accuracy of the vertical field.
Furthermore, high resolution
vector magnetographs such as Hinode/SOT have a limited field of view,
which often does not allow deriving the horizontal
magnetic field in an entire
active region. The influence of these effects on the accuracy of non-linear
force-free extrapolations have recently been investigated in \cite{derosa:etal09}.
Less often considered are
effects introduced by the fact that the extraction of the field from the
Stokes parameters is intrinsically uncertain. E.g., the measured Stokes
profiles are formed in a highly dynamic atmosphere with a complex thermal and
magnetic structure, while the normally applied inversion methods
 impose rather restrictive assumptions
 \citep[e.g. Milne-Eddington atmosphere,][]{auer:etal77}.
 In addition, due to the spatially fluctuating height of formation
of the lines, the obtained magnetic vector refers to different heights in the
atmosphere at different locations. For extrapolation methods one has to
assume a single, often planar, height as the boundary condition.
Finally, instrumental limitations impose
restrictions, such as limited spatial and spectral resolution, spectral
sampling (in particular for filter instruments) and noise.

Here we
investigate how the extrapolated coronal magnetic field is influenced by
noise and other instrumental artifacts (spatial resolution,
limited spectral sampling), as well as the general limitations of the
inversion of the measured Stokes profiles.
To do so we use the results of 3D radiation MHD simulations. We compute
synthetic lines from the data cubes of the relevant physical quantities, add
noise and apply the influence of typical instrument parameters. Finally we
invert these artificial measurements to derive synthetic vector magnetograms,
which are then used as boundary conditions for a nonlinear force-free
magnetic field extrapolation. We compare the reconstruction from ideal data,
taken directly from the MHD simulations, with extrapolations starting from
data containing instrumental effects and noise of different levels of
severity. Our aim is to investigate how different instrumental effects and
noise influence a nonlinear force-free magnetic field extrapolation. Besides
testing generally the influence of reduced spatial resolution of the
photospheric magnetic field data, we also consider to what extend the
properties of specific high resolution space instruments affect the quality
of the extrapolations. The two instruments we consider are the
Spectro-Polarimeter (SP) on the Solar Optical Telescope (SOT) on the Hinode
spacecraft \citep[][]{shimizu04, 2008SoPh..249..167T}
and the Polarimetric and Helioseismic Imager
(PHI) to fly on the Solar Orbiter mission. Instruments with lower spatial
resolution, such as the Helioseismic Magnetic Imager (HMI) were not
considered, since the number of pixels across the MHD simulation box are then
rather small, making a meaningful test more improbable. %
\section{Method}
\subsection{Setup of the test case}
\label{sec2.1}
We start with 3-D radiation MHD simulations resulting from the MURAM code
\citep[see][for details]{voegler:etal05}. The particular snapshot considered
here has been taken from a bipolar run and harbours equal amounts of magnetic
flux of both polarities. The configuration has an average field strength of
$150$ G at the spatially averaged continuum optical depth $\langle \tau_{5000} \rangle =1$
at $5000$ {\AA}. The horizontal dimensions of the simulation domain are $6
\times 6 {\rm Mm}$ and $1.4 {\rm Mm}$ in the vertical direction. The original
resolution is $20 {\rm km}$ in the horizontal directions and $14 {\rm km}$ in
the vertical direction. Similar snapshots have also been used by
\cite{khomenko:etal05a} for the investigation of magnetoconvection of
mixed-polarity quiet-Sun regions and compared with measured Stokes profiles
by \cite{khomenko:etal05}.

\begin{figure*}
   \centering
\includegraphics[clip, bb=0 65 752 290, width=14cm,height=4cm]{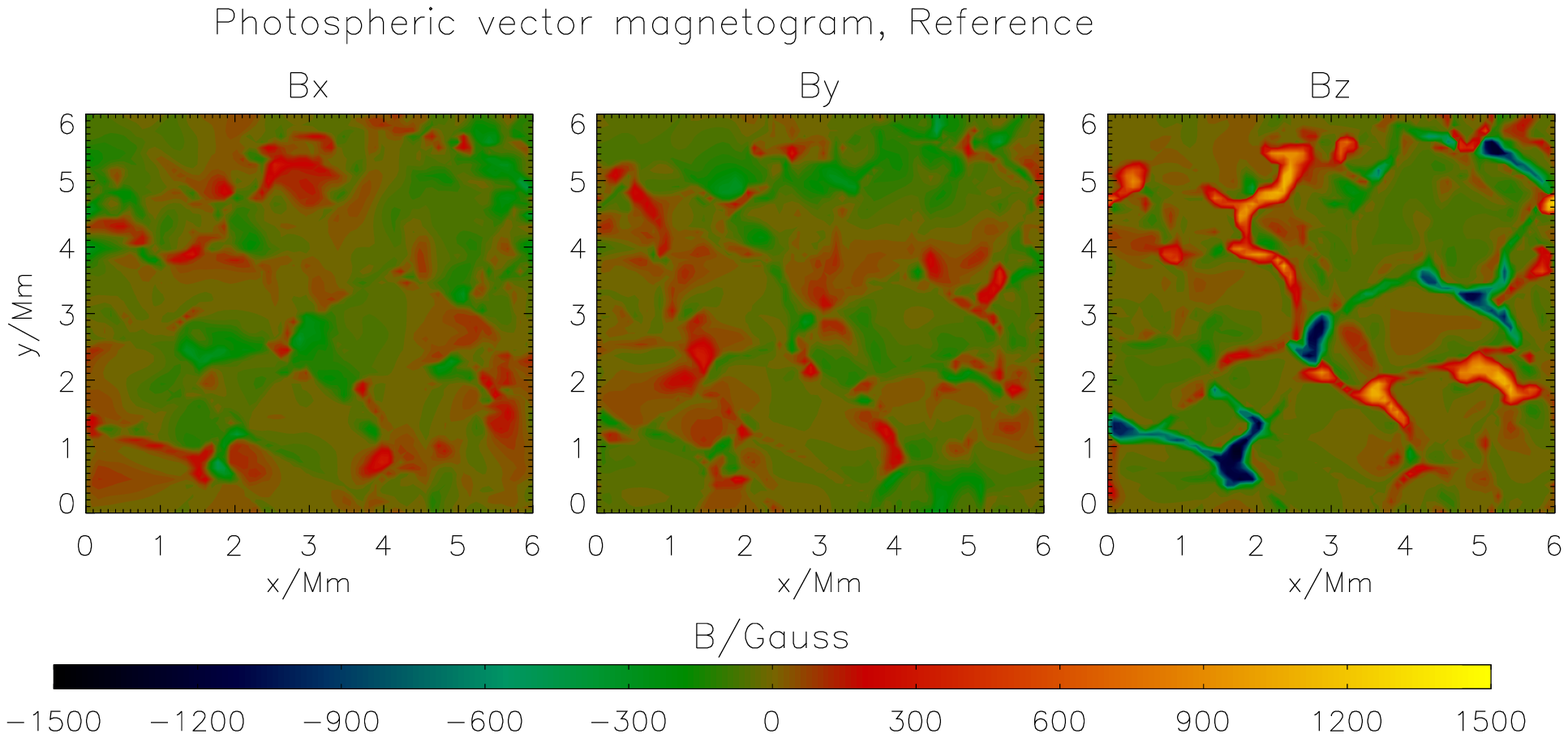}
\includegraphics[clip, bb=0 65 752 290, width=14cm,height=4cm]{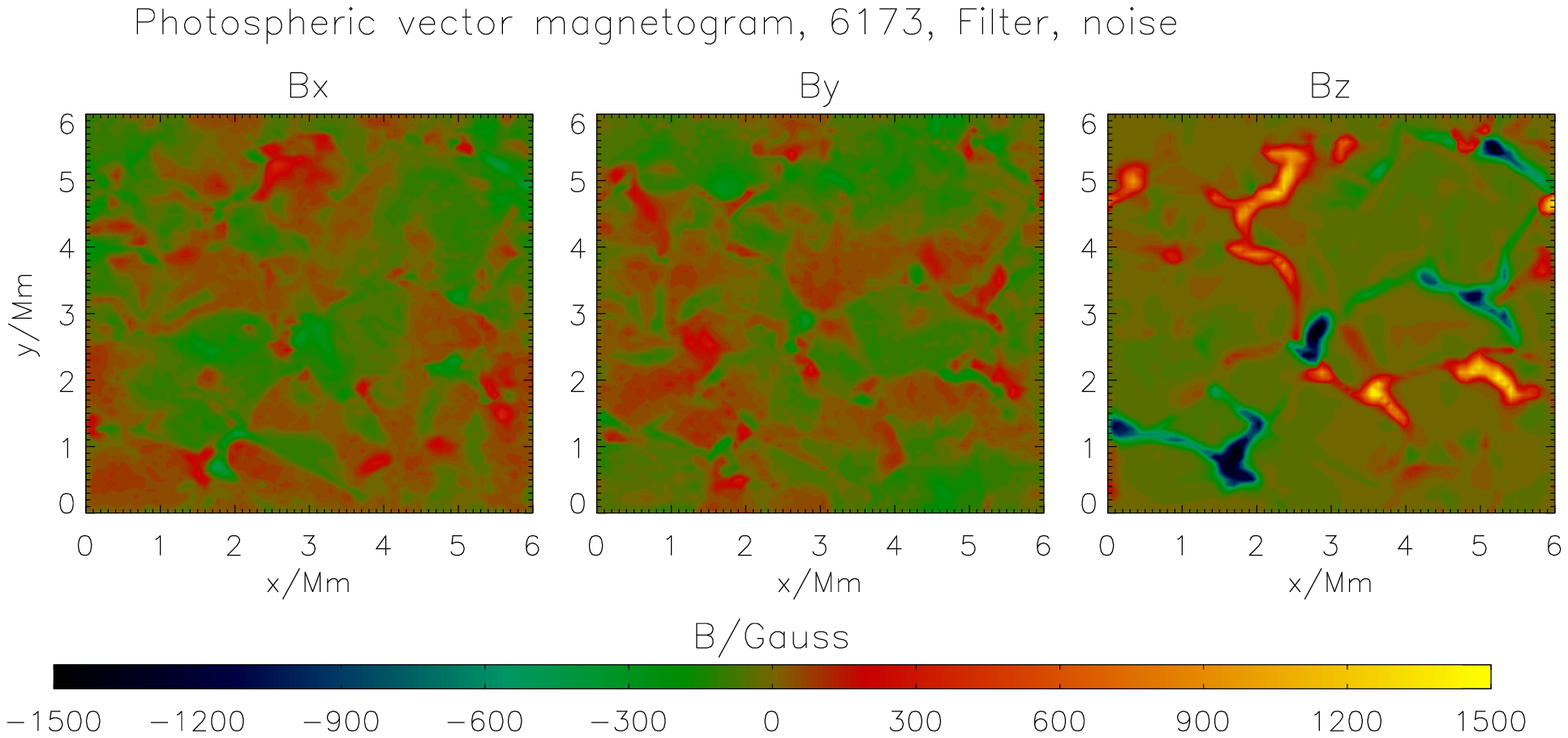}
\includegraphics[clip, bb=0 65 752 290, width=14cm,height=4cm]{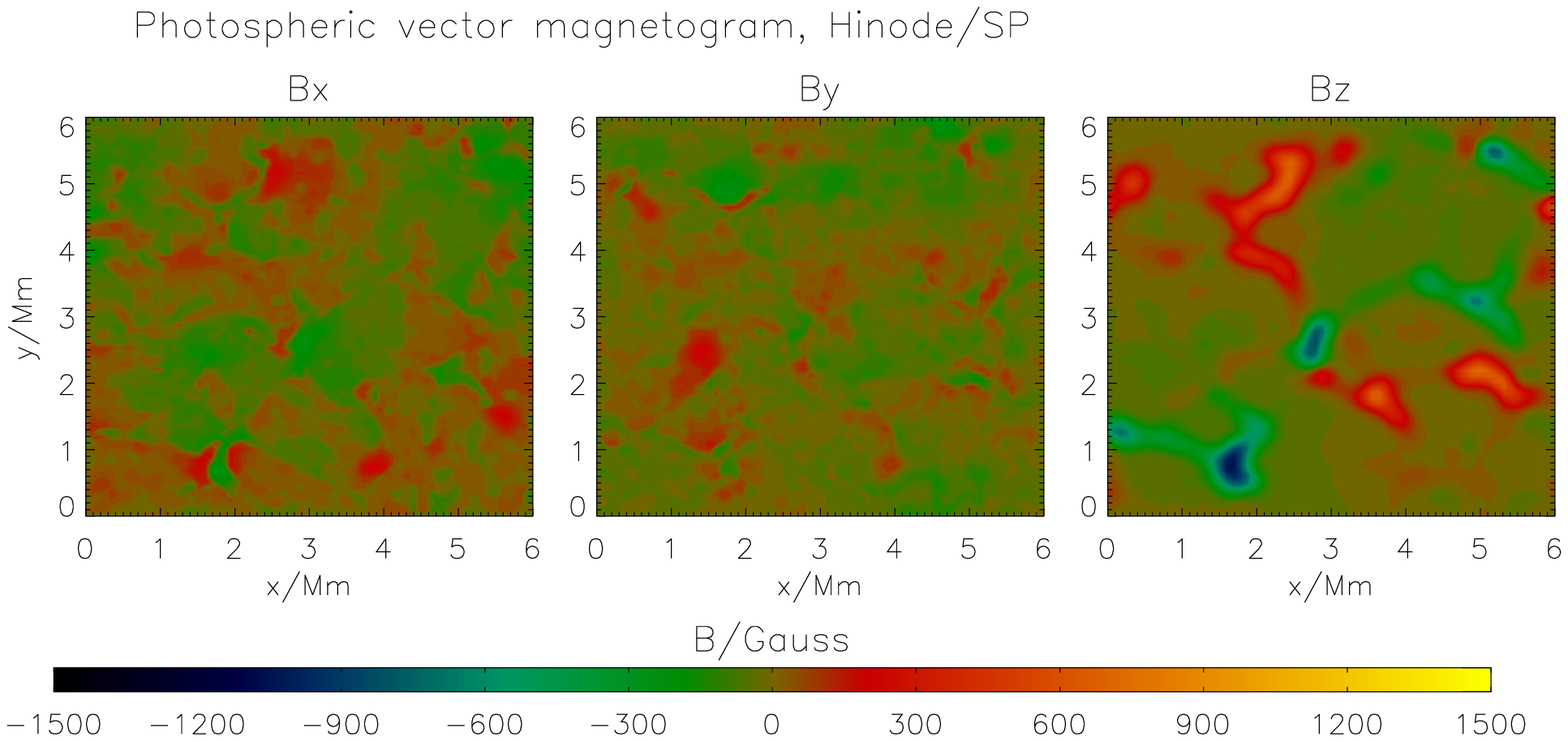}
\includegraphics[clip, bb=0 0 752 290, width=14cm,height=5cm]{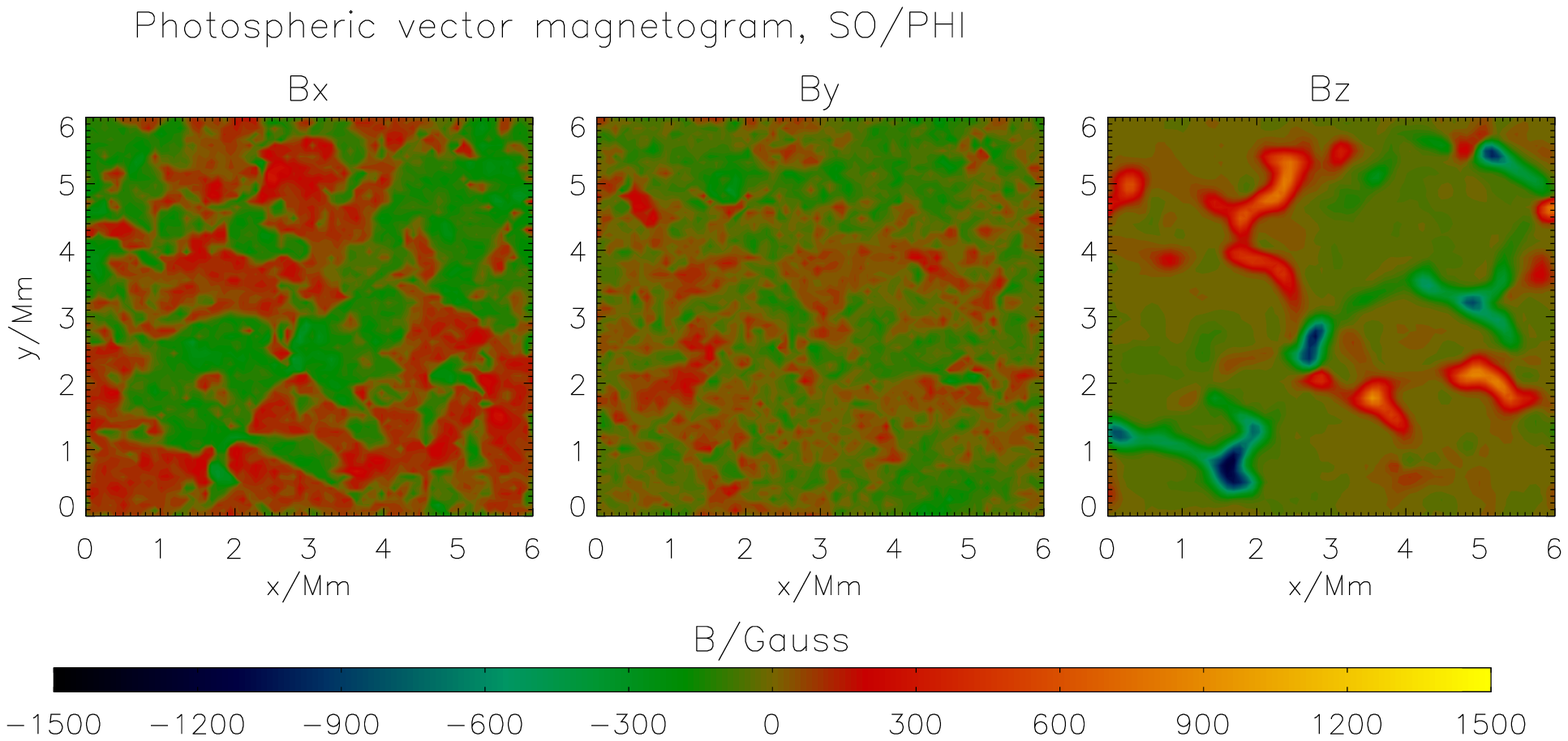}
\caption{$B_x, B_y, B_z$ maps at a fixed reference height
$Z_{\rm ref}$ approximately 150 km above the average $< \tau_{5000} >= 1$.
The top panel contains the original maps from the MHD data, which are used
as a reference. The second panel shows the results of an inversion with
Filter and noise. The third and bottom panel correspond to simulated inversions
from Hinode/SP and SO/PHI, respectively. For a quantitative comparison see
Table \ref{table1}.}
\label{fig0}
\end{figure*}

Due to the limited height-range of the MHD simulation we
first extrapolate the field from the magnetic vector
obtained from the simulations at a fixed geometric height
(see Fig. \ref{fig0}). This
height, $Z_{\rm ref}$, must be chosen with care since the extrapolation
starting from the magnetic field vector at  $Z_{\rm ref}$ is employed as a
reference against which all others are compared. In order to avoid
introducing a bias it must correspond to a suitably weighted average
height of formation of the spectral line used for inversion, e.g.
Fe I $6173.3$ {\AA}. Since different regions in the MHD-simulation may have
different formation height ranges, we have chosen a criterion to
select the reference height which is obtained by finding the highest
correlation between the magnetic field strength taken directly from
the MHD simulation and from inversion of the Stokes profiles
obtained under ideal conditions (very high spectral and spatial
resolutions and no noise).

This extrapolation is used as reference, with which other
extrapolations are compared. These later ones use magnetic vector
maps that are obtained from the inversion of synthetic Stokes
profiles.

Two spectral lines are considered, the very widely used Fe I $6302.5$
{\AA} (Land\'{e} factor $g=2.5$) line, employed by the Advanced Solar
Polarimeter and the spectropolarimeter on the Hinode spacecraft
and the  Fe I $6173.3$ {\AA} ($g=2.5$), which has been selected for the
Helioseismic and Magnetic Imager (HMI) on the Solar Dynamics
Observatory (SDO) and for the Polarimetric and Helioseismic Imager
on the Solar Orbiter mission (SO/PHI). For both $Z_{\rm ref}$ was
found to lie approximately at $150$ km above $ \langle \tau_{5000} \rangle =1$.

Next, using the STOPRO code \citep{solanki1987, frutiger:etal00}, we compute
the Stokes parameters of these widely used spectral lines. We restrict
ourselves to considering the situation at solar disk center. These
synthetic data are then either directly inverted, or after degrading the data
by introducing noise, lowering the spatial and/or the spectral resolution,
etc. A list of all the considered cases is presented in Table 1 and discussed
in Sect. 3. The inversion is carried out for each spectral line
individually. The inverted magnetic field vector is then taken as the
starting point for a non-linear force-free extrapolation.

For comparison we also compute a potential field from the results
of the MHD-simulations at $Z_{\rm ref}$, which requires only the
vertical component of the photospheric magnetic field. In
addition, we compute nonlinear force-free extrapolations with the
horizontal spatial resolution reduced by a factor of two and four,
respectively, compared to the original MHD model.
\subsection{Description of the inversion code}

The inversion code {\bf HeLIx} developed by \cite{lagg:etal04} is based on
fitting the observed Stokes profiles with synthetic ones obtained from an
analytic solution of the Unno-Rachkovsky equations
\citep{unno:56,rachkowsky:67} in a Milne-Eddington atmosphere \citep[see
e.g.][]{landi92}. These synthetic profiles are functions of the magnetic field
strength $|B|$, its inclination and azimuth (with respect to the
line of sight), the vertical velocity, the
Doppler width, the damping constant, the ratio of the line center to the
continuum opacity, and the slope of the source function with respect to
optical depth. We assume the filling factor to be unity. Any possible
  magnetic fine structure within one pixel consisting of a field-free and a
  magnetic area is smeared out, and the inversion returns the magnetic field
  vector averaged over the considered pixel. Since the extrapolations are
  performed on the same pixel scale as the inversions, the extrapolations must
  be fed with the pixel-averaged magnetic field vector. Setting the filling
  factor to unity already during the inversion process enhances its robustness
  and reliability. The atmospheric parameters that ensure the minimum of the
merit function are obtained using a very reliable genetic algorithm
\citep{charbonneau95}.   The genetic algorithm has been extensively tested
  with synthetic spectra from MHD simulations and the result compared with
  response-function-averaged physical parameters (e.g. magnetic field
  strength, inclination, azimuth, line-of-sight velocity). The results of the
  test
  indicate that the genetic algorithm retrieves the global minimum of the
  merit function with high reliability. The parameter sets are chosen to,
e.g., mimic the effects of the Hinode/SOT and the future SO/PHI instruments,
respectively.
\subsection{Hinode SOT Spectro-polarimeter}
\label{hinode_cases} The spectropolarimeter
\citep[SP;][]{lites:etal07} is part of the focal-plane package of the $50
cm$ Solar Optical Telescope (SOT) onboard the Hinode spacecraft. It
observes the line pair Fe I $6302.5$ and $6301.5$ {\AA}. Here we
restrict ourselves to $6302.5$ {\AA}. The spatial resolution at the
diffraction limit of the telescope's primary mirror is about 0.32"
at $6302.5$ {\AA}, which corresponds to 230 km on the Sun.
 The size of a detector pixel corresponds to approximately 110 km
on the Sun in the spatial direction.
The spectral resolution is $30$ m{\AA} and the spectral sampling is $21$
m{\AA}. We used Gaussians for spectral and spatial smearing.
 The noise has been added to the Stokes profiles as photon noise
$F \times 1/ \sqrt{I_c}$, where $F$ is a white noise with a Gaussian distribution
and $I_c$ is the continuum intensity.
The chosen standard deviation of  $F$ was $10^{-3}$, which corresponds to the typical noise
level for modern spectropolarimetric observations (e.g. Hinode/SP data).
Noise is included in a similar way wherever we mention that noise has been added to the data.
More subtle instrumental effects such as scattered light, or a slight
defocus \citep[e.g.][]{danilovic:etal08}, are not considered.
\subsection{SO/PHI magnetograph}
\label{PHI_cases}
The Solar Orbiter Polarimetric and Helioseismic Imager (SO/PHI), a vector
magnetograph, will be one of the main instruments on the ESA-NASA
Solar Orbiter mission. Of its two telescopes, the High Resolution
Telescope (HRT) is of primary importance for the present
investigation (due to the small horizontal extent of the MHD
simulations). The spectral line chosen
for SO/PHI, Fe I $6173.3$ {\AA},
combines high Zeeman sensitivity with spectral purity, needed for
simultaneous vector magnetic field and helioseismology studies. We
describe the point spread function (PSF) by a Gaussian with
FWHM$=150$ km. The obtained arrays of Stokes parameters are rebinned
to a spatial pixel size of $80$km. We then convolve the Stokes
profiles in the spectral dimension with a  Fabry-P\'erot type filter
with ${\rm FWHM}=100$ m{\AA}. Since SO/PHI will be a filtergraphic
instrument, we decrease the number of spectral samples per line by
taking 5 positions in the line and one in the continuum at the
positions (from line center at rest): -0.3{\AA}, -0.15{\AA}, -0.075
{\AA}, 0 {\AA}, +0.075 {\AA},   +0.15 {\AA}. At that stage we add
noise and perform the inversion of the Stokes profiles.
\subsection{Extrapolation of the vector magnetogram into the atmosphere}
To compute a 3D-nonlinear force-free magnetic field from the result
of the  {\bf HeLIx} inversion code we carry out the following steps:
\begin{itemize}
\item If needed, transform  $B, \, \theta$ and $\phi$, which are output from the
inversion code, to $B_x, \, B_y, \, B_z$ on the photosphere, which requires a
resolution of the $180^{\circ}$ ambiguity in $\phi$. (See section
\ref{sec2.5.1}.)
\item Preprocess the vector magnetogram  ($B_x, \, B_y, \, B_z$), assuming that
it refers to the same geometric height at every spatial pixel. (See
Sect. \ref{sec2.5.2}.)
\item Compute a nonlinear force-free coronal magnetic field from the preprocessed
 vector magnetogram.
(See Sect. \ref{sec2.5.3}.)
\item Compare the result with the reference field.
\end{itemize}
We explain these steps in the following.
\subsubsection{Removal of the $180^{\circ}$ ambiguity.}
\label{sec2.5.1}
 For the purpose of the present investigation we have chosen to remove
 the $180^o$ ambiguity by minimizing the angle to the exact solution.
   This possibility does, of course, not exist for real data and
   other methods for the ambiguity inversion would need to be tried.
 The performance of different ambiguity removal techniques has  been studied with
   synthetic data by \cite{metcalf:etal06}. They found that the best available
   technique managed to get $ 100 \%$ of the points correctly.
   Recently the influence of noise and spatial resolution
   on the quality of the different ambiguity removal techniques
   has been investigated by \cite{leka:etal09}.
   We have therefore not
   considered specific ambiguity removal techniques. An investigation of their
   efficiency and influence is outside the scope of this paper.
\begin{table}
\caption{Influence on the photospheric field for different cases.}
 \label{table1}
 \centering
\begin{tabular}{ccccccc}
\hline
Case studied    & $C_{Bx}$&$C_{By}$&$C_{Bz}$&$C_{\rm vec}$&$<|J_z|>$ &$<|B_z|>$ \\
\hline
{\bf Reference} & $ 1$ & $1$ &$1$ & $1$& $0.059$ & $103.5$ \\
\multicolumn{4}{l}{\bf Inversion of synthetic profiles (tests)} &&&\\
\multicolumn{4}{l}{Full resolution, no noise and full profiles} &&&\\
$6302.5$ {\AA} & $ 0.89$ & $0.88$ &$0.97$ & $0.96$& $0.065$ & $96.7$  \\
$6173.3$ {\AA} & $ 0.89$ & $0.88$ &$0.97$ & $0.96$& $0.065$  & $95.9$ \\
\multicolumn{4}{l}{Full resolution, with noise and full profiles} &&\\
$6302.5$ {\AA} & $ 0.87$ & $0.86$ &$0.97$ & $0.95$& $0.086$ &$97.6$  \\
$6173.3$ {\AA} & $ 0.88$ & $0.88$ &$0.97$ & $0.95$& $0.070$ &$96.5$ \\
\multicolumn{4}{l}{Full resolution, with Filters $(5+1 \lambda)$ values} &&\\
$6173.3$ {\AA} no noise & $ 0.86$ & $0.85$ &$0.97$ & $0.95$& $0.084$ & $96.0$\\
$6173.3$ {\AA} w. noise & $ 0.72$ & $0.70$ &$0.97$ & $0.90$& $0.213$ & $96.7$\\
\hline
{\bf Hinode/SP}  &&&&&\\
no noise  & $ 0.80$ & $0.77$ &$0.84$ & $0.82$& $0.047$& $79.4$\\
$10^{-3}$ noise  & $ 0.76$ & $0.71$ &$0.84$ & $0.81$& $0.067$& $80.7$\\
{\bf SO/PHI}  &&&&&\\
 no noise & $ 0.79$ & $0.68$ &$0.85$ & $0.82$& $0.078$& $84.5$\\
 $10^{-3}$ noise & $ 0.67$ & $0.53$ &$0.85$ & $0.73$& $0.184$&$85.2$\\
\hline
\end{tabular}
\end{table}
%
\subsection{Effects in the photosphere}
In Table \ref{table1} and in figure \ref{fig0}we investigate how the different
instrument effects and noise influence the vector field in the photosphere.
The first line corresponds to the MHD reference field. The field has an
average electric current density of $\langle |J_z| \rangle =0.059 A/m^2$ and a vertical
magnetic field strength of $ \langle |B_z| \rangle =103.5 G$. Positive and negative
values of these quantities are balanced.
In the table we compute
the correlation relative to this reference case for the horizontal fields
$C_{Bx}, \, C_{By}$, the vertical field $C_{Bz}$ and the 2D-vector correlation
in the photosphere $C_{\rm vec}$ and provide the average absolute values
$\langle |B_z| \rangle $ and $\langle |J_z| \rangle$.
For all full spatial resolution
cases (upper part of the table) the correlation for the different cases is
$0.97$ for the vertical magnetic field strength and the average field strength
is underestimated by a few percent. Noise and instrument effects seem to have
a relative small effect on the vertical field. The effect of a reduced resolution
(lower part of the figure, Hinode/SP and SO/PHI cases)
 on the horizontal photospheric magnetic fields $B_x, \, B_y$ and
the derived vertical current density $J_z$  is significantly higher. While for
full profiles the correlation in the horizontal fields is in the range of $0.85-0.88$,
the combined effect of filter and noise reduces the correlation to only $0.70$ and
spurious, non-physical electric currents, which results in an overestimation
of $J_z$ by a factor of about $3$.
\subsubsection{Preprocessing}
\label{sec2.5.2}
 The magnetic field in the photosphere is not necessarily force-free
 \citep[because of the finite $\beta$ plasma in the photosphere, see][]{gary01}
 and the horizontal components
 ($B_x$ and $B_y$) of current  vector magnetographs have large uncertainties.
 \citet{aly89} defined a number of integral relations
 to evaluate if a measured photospheric vector magnetogram is consistent with the
 assumption of a force-free field. These integral relations
 (numerator in Eq. \ref{aly1}) have been used to define
 a dimensionless parameter $\epsilon_{\mbox{force}}$ as
 \begin{equation}
\frac{|\int_{S} B_x B_z \;dx\,dy| + |\int_{S} B_y B_z \;dx\,dy|+
|\int_{S} (B_x^2+B_y^2)-B_z^2 \;dx\,dy |}
{\int_{S} (B_x^2+B_y^2+B_z^2) \;dx\,dy}
\label{aly1}
 \end{equation}
and force-free extrapolation codes require $\epsilon_{\mbox{force}} \ll 1$ on
the boundary. (One gets $\epsilon_{\mbox{force}} =0$ if the integral relations are
fulfilled exactly).

 For the synthetic magnetograms investigated in table \ref{table1} we find
 $\epsilon_{\mbox{force}}=0.57 \pm 0.19$ in the photosphere.
\citet{wiegelmann:etal06} developed a preprocessing procedure to drive the
 observed non force-free data
 towards  boundary conditions suitable for a force-free extrapolation.
 As a result of the preprocessing we get a boundary-data set which is consistent
 with the assumption of a force-free magnetic field.
After applying the preprocessing-routine we get force-free consistent boundary
conditions with
$\epsilon_{\mbox{force}}= (2.6 \pm 0.2) \cdot 10^{-4}$.
The preprocessing affects mainly the horizontal magnetic field components.
The correlation of original and preprocessed field for the investigated cases is
$C_{Bz}=0.99 \pm 0.006$ for the vertical component $B_z$.
For the horizontal components we find a correlation of
$C_{Bx}=0.91 \pm 0.05$ and $C_{By}=0.90 \pm 0.05$, respectively.

\subsubsection{Extrapolation of nonlinear force-free fields.}
\label{sec2.5.3}
 \label{basic}
Force-free coronal magnetic fields have to obey the equations
\begin{eqnarray}
(\nabla \times {\bf B })\times{\bf B} & = & {\bf 0},  \label{forcefree}\\
\nabla\cdot{\bf B}    & = &         0      \label{solenoidal-ff}.
\end{eqnarray}
We define the functional
\begin{equation}
L=\int_{V} \;  w \; \left[ \; B^{-2} \, |(\nabla \times {\bf B})
\times {\bf B}|^2 +\; |\nabla \cdot {\bf B}|^2\right] \; d^3x
\label{defL1},
\end{equation}
where $w$ is a weighting function. It is obvious that (for $w>0$) the
force-free equations (\ref{forcefree}-\ref{solenoidal-ff}) are fulfilled when
L is equal to zero \citep[][]{wheatland:etal00}. We minimize the functional
(\ref{defL1}) numerically as explained in detail by \cite{wiegelmann04}. The
program is written in C and has been parallelized with OpenMP.
\citet{wiegelmann:etal03} and \citet{schrijver:etal06} tested the program
with exact nonlinear force-free equilibria of \cite{low:etal90}, while
\cite{wiegelmann:etal06a} tested it with another exact equilibrium developed
by \cite{titov:etal99}. The code has been applied to vector magnetograph data
from the German Vacuum Tower Telescope (VTT) by
\cite{wiegelmann:etal05b,wiegelmann:etal05} and to data from the Solar Flare
Telescope (SFT) by \cite{wiegelmann:etal06}. Here we use an updated version
of the optimization approach, including a multi-scale approach which has been described
and tested by \cite{metcalf:etal08} and applied to Hinode data by
\citep{schrijver:etal08}.
%
\begin{table}
\caption{Various figures of merit of the 3D reconstruction for different
cases.}
 \label{table2}
 \centering
\begin{tabular}{cccccc}
\hline
   Case studied    & $C_{\rm vec}$&$\epsilon$&$C_{\rm 100km}$&$C_{\rm 400km}$&$C_{\rm 800km}$ \\
      \hline
{\bf Reference} & $ 1$ & $1$ &$1$ & $1$& $1$\\
Potential & $0.68$ &$0.62$ &$0.72$ &$0.84$& $0.97$ \\
\multicolumn{4}{l}{\bf MHD cases with reduced resolution} &&\\
Pixel Size 40km & $0.99$&$0.93$&$0.99$&$1.00$ & $1.00$\\
Pixel Size 80km & $0.93$&$0.79$ &$0.94$&$0.96$ & $0.99$\\
\hline
\multicolumn{4}{l}{\bf Inversion of synthetic profiles (tests)} &&\\
\multicolumn{4}{l}{Full resolution, no noise and full profiles} &&\\
$6302.5$ {\AA} & $0.92$ &$1.06$&$0.93$ &$0.98$& $1.00$ \\
$6173.3$ {\AA} &$0.93$&$0.99$&$0.93$&$0.98$& $0.99$\\
\multicolumn{4}{l}{Full resolution, with noise and full profiles} &&\\
$6302.5$ {\AA} & $0.93$ &$1.04$&$0.93$ &$0.98$& $1.00$ \\
$6173.3$ {\AA} & $0.93$ &$1.05$&$0.93$ &$0.98$& $0.99$ \\
\multicolumn{4}{l}{Full resolution, with Filters $(5+1 \lambda)$ values} &&\\
$6173.3$ {\AA} no noise & $0.92$ &$1.06$&$0.92$ &$0.98$& $0.99$\\
$6173.3$ {\AA} w. noise & $0.91$ &$1.10$&$0.89$ &$0.97$& $0.99$\\
\hline
{\bf Hinode/SP}  &&&&&\\
no noise  &$0.85$&$0.66$&$0.85$&$0.96$&$0.99$\\
$10^{-3}$ noise  &$0.84$&$0.62$&$0.84$&$0.95$&$0.99$\\
{\bf SO/PHI}  &&&&&\\
 no noise &$0.84$ &$0.75$&$0.85$ &$0.96$& $0.99$ \\
 $10^{-3}$ noise &$0.81$ &$0.76$&$0.78$ &$0.90$& $0.98$ \\
\hline
\end{tabular}
\end{table}
%
\begin{figure*}
   \centering
\includegraphics[width=18cm]{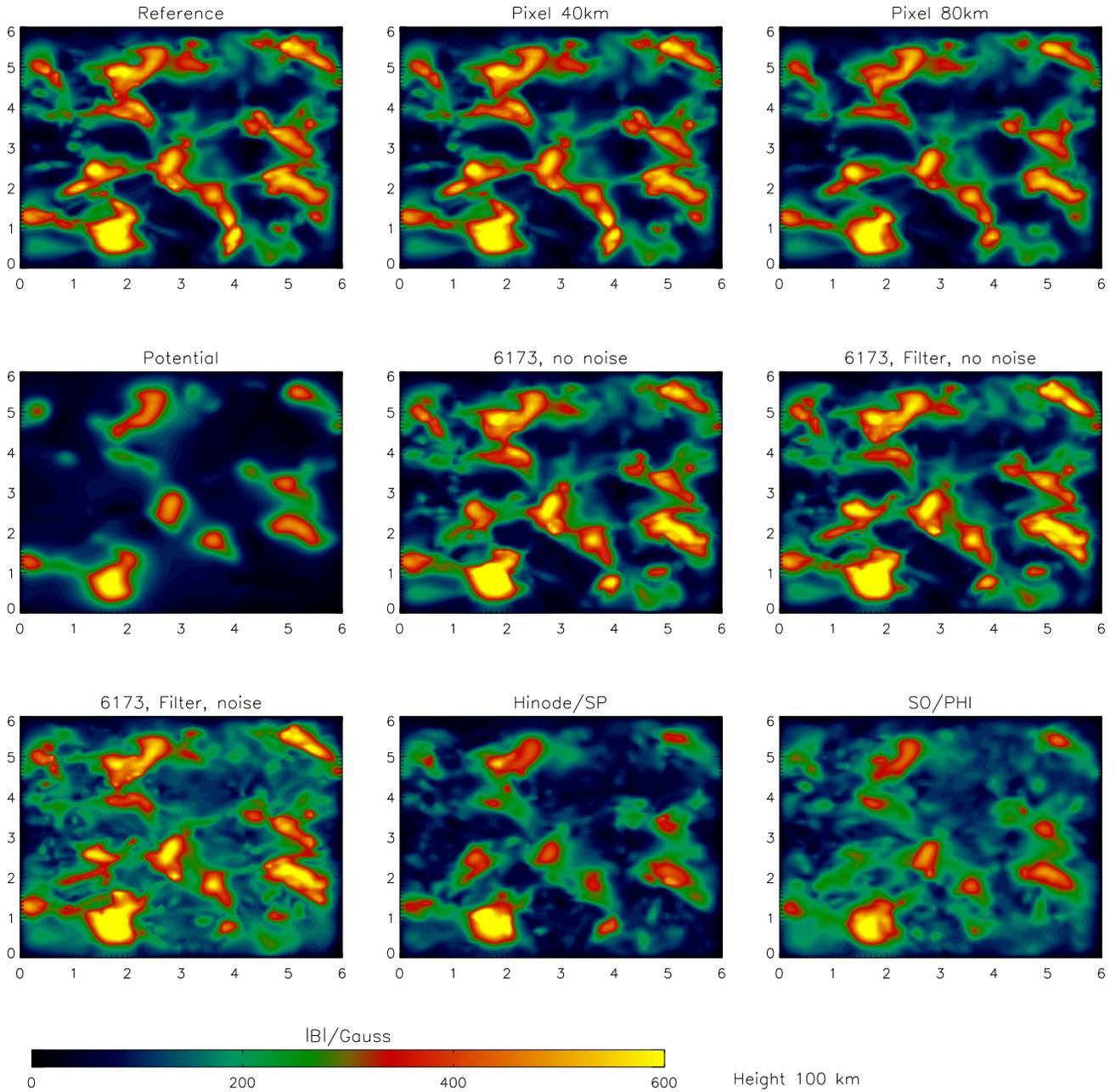}
\caption{$|{\bf B}|$ for the reference field and extrapolated from inversion
of synthetic profiles. The plots display $|{\bf B}|$ at a height of $100 km$
above the reference height (See Sect. \ref{sec2.1} for a definition). The
figure shows, from top-left to bottom-right, the ideal reference solution
(indicated as {\bf Reference}); extrapolations starting from the MHD output
after reducing its spatial resolution ({\bf Pixel 40km} and {\bf Pixel 80km},
respectively); a potential field reconstruction ({\bf Potential}); a
reconstruction from a full resolution spectropolarimeter in the $6173.3$ {\AA}
line without noise and inverting the full profiles ({\bf 6173, no noise});
reconstructions starting from inversions of full resolution filter
magnetograms without and with noise ({\bf Filter, no noise} and {\bf Filter, noise},
respectively); and finally two cases with specifications adapted to those of
two space-borne instruments, the spectropolarimeter on Hinode ({\bf
Hinode/SP}) and the Polarimetric and Helioseismic Imager on Solar Orbiter
({\bf SO/PHI}). x and y axis are in Mm.}
\label{fig2}%
\end{figure*}
\begin{figure*}
   \centering
\includegraphics[width=18cm]{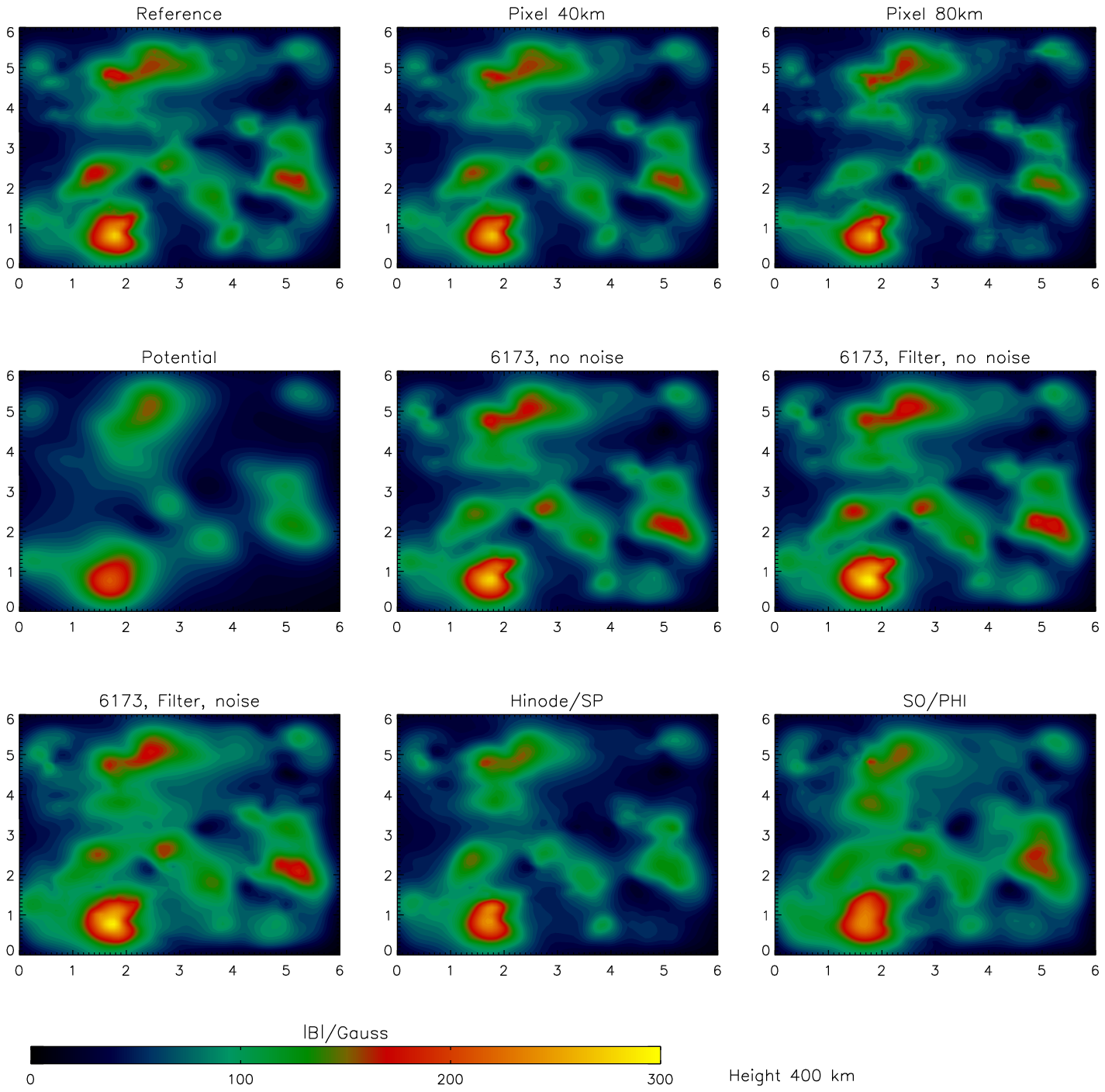}
\caption{Same as Fig. \ref{fig2}, but $400 km$ above the reference height.}
\label{fig2b}%
\end{figure*}
\subsection{Figures of merit}
\cite{schrijver:etal06} introduced several figures of merit to
compare the results of magnetic field extrapolation codes
(a 3D-vector field ${\bf b}$) with a reference solution ${\bf B}$.
\begin{itemize}
\item Vector correlation:
\begin{equation}
C_{\rm vec}= \sum_i {\bf B_i} \cdot {\bf b_i}/ \left( \sum_i |{\bf B_i}|^2
\sum_i |{\bf b_i}|^2 \right)^{1/2},
\end{equation}
where $i$ corresponds to all grid points in the entire 3D computational box.
\item Total magnetic energy of the reconstructed field
${\bf b}$ normalized by the energy of the reference field ${\bf B}$:
\begin{equation}
\epsilon = \frac{\sum_i |{\bf b_i}|^2}{\sum_i |{\bf B_i}|^2}.
\end{equation}
\item We also compute the linear Pearson correlation $C$ of the total magnetic field
strength $|B|$ at the heights 100km, 400km and 800km above the reference
height, $Z_{ref}$, respectively.
 \end{itemize}
Two vector fields agree perfectly if $C_{\rm vec}$, $\epsilon$ and the
Pearson correlation coefficients  are unity.
\section{Results}
Table \ref{table2} contains the different quantitative measures
 which compare the different reconstructed field from
 synthetic observations with the
 reference field (extrapolations from ideal data). Column 2 contains the
 vector correlation $C_{\rm vec}$, column 3 the normalized magnetic energy
 $\epsilon$, and columns 4-6 the linear Pearson correlation $C$ of the total magnetic field
strength $|B|$ at the heights 100km, 400km and 800km above the reference
height, $Z_{ref}$, respectively.

Figs \ref{fig2} and \ref{fig2b} show contourplots of the magnetic field
strength $|{\bf B}|$ at $100$ km and $400$ km above $Z_{ref}$ respectively.
\subsection{MHD cases}
In the first $4$ images of Figs \ref{fig2} and \ref{fig2b} we compare the
nonlinear force-free 3D magnetic field reconstructed from ideal data
(extracted from the MHD simulations at $Z_{ref}$, called {\bf Reference} in
Table \ref{table2} and in Figs. \ref{fig2}, \ref{fig2b})
 with a potential field extrapolation also starting from ideal data ({\bf Potential}) and
nonlinear force-free computations starting from magnetic field maps obtained
 from the inversion of synthetic Stokes profiles and MHD cases.
Potential fields are often calculated in
addition to NLFFF, because they contain the minimum energy for given vertical boundary
conditions and the free energy of a NLFF-field above that of a potential field has relevance
for coronal eruptions.

As seen in Table \ref{table2} and in Figs. \ref{fig2} and \ref{fig2b}
the computations with reduced spatial resolution of $40$ km pixels
show almost perfect agreement, while $80$ km pixels still provide an
excellent correspondence. The correlation with the reference field is better
higher in the atmosphere, which is not surprising since small-scale field
structures at the lower boundary do not propagate very high and the larger
scales are less affected by the binning to larger horizontal pixels. The
magnetic energy is underestimated because small scale fields and currents at
low heights, which contribute significantly to the total magnetic energy, are
not resolved here.

\subsection{Dependence on spectral line, noise and spectral sampling}
The remaining images in Figs. \ref{fig2} and \ref{fig2b} and the remaining
entries in Table \ref{table2} apply to extrapolations starting from magnetic
maps obtained from inversions of synthetic spectral lines. The first $3$ of
these images (marked 6173 no noise, Filter no noise, Filter noise in Figs. \ref{fig2}
and \ref{fig2b}) and the whole middle part of Table \ref{table2} are test
cases for which we move step-by-step away from ideal conditions (i.e.
$\mathbf{B}$ obtained directly from the MHD simulations) towards more
realism, i.e. introducing various instrumental effects.

The first such step is to invert the synthetic line profiles directly,
without any further manipulation (ideal spectropolarimeter). This step
introduces uncertainties due to the inversion process per se
(e.g. by the fact that asymmetric Stokes profiles are fit assuming
purely symmetric or antisymmetric ones), the fact that
the line samples a range of heights, and the dependence of the line formation
height on the type of solar feature.

The inversion of Stokes spectra expected from an ideal spectropolarimeter
leads to an agreement with the reference within a few percent for the vector
correlation and we can estimate the magnetic energy with an accuracy of one
percent for the $6173.3$ {\AA} line. The magnetic energy is an important
quantity because it tells us how much free energy is maximally available for
eruptive phenomena like flares and coronal mass ejections. The high accuracy
achieved in this case is particularly encouraging since it shows that
applying a simple Milne-Eddington inversion to the often highly asymmetric
profiles (sometimes showing multiple lobes in Stokes V) gives sufficiently
accurate results and that the fluctuating height at which the magnetic field
is obtained does not significantly influence the extrapolations (note that
the situation may be different in a sunspot with its rather deep Wilson
depression). This result implies that most of the inaccuracies in the
extrapolations are due to limitations in the instrumentation and not because
it is in principle not possible to extract the information from the
observations. %

As Table \ref{table2} shows there is little to choose between the two $g=2.5$
Zeeman triplets $\lambda 6302.5$ {\AA} line and $\lambda 6173.3$ {\AA}. The field
extrapolated from the magnetic field maps derived from either of them
correlate very well with the reference field. Using the $\lambda 6173.3$ {\AA}
lines leads to a slightly better estimate of the magnetic energy, but all in
all the necessity of using the Zeemann effect in a spectral line by itself
only leads to errors of a few percent \footnote{The $Z_{\rm ref}$
was set to the average formation height for the $\lambda 6173.3$ {\AA} line. This might explain the
somewhat worse results for the $\lambda 6302.5$ {\AA} line.}. The addition of
noise at a level of $10^{-3}$ of the continuum intensity $I_c$, which is typical of modern
spectropolarimetric observations, has only a small effect. Note that by
adding an equal amount of photon noise to all Stokes parameters we are
producing a much lower S/N ratio in the linearly polarized $Q$ and $U$
profiles than in Stokes $V$ due to the smaller signal in the former.  The influence
of noise of a given amplitude also depends heavily on the
magnetic flux in the box for which we extrapolate. With $\langle B \rangle
\approx 150$ G the chosen snapshot corresponds to an average plage region. We
expect that the same amount of noise will have a considerably stronger effect
when applied to the weaker Stokes profiles present in the quiet sun.

It is of particular interest if measurements in limited wavelength bands and at
reduced wavelength resolution, typical of filter polarimeters (filter
magnetographs), are acceptable for extrapolations. One advantage of such
instruments is that they allow time series of a whole region to be recorded
at high cadence. Of particular relevance for magnetic extrapolations is the
fact that filter polarimeters
 record the Stokes vector over the full field of view. This
overcomes the main shortcoming of spectropolarimeters, namely that they need
to scan a region step by step, so that by the time the second footpoint of a
loop is scanned the first may have evolved considerably. These advantages of
filter instruments come at the price of a reduced spectral resolution and
limited spectral sampling. An extensive series of tests by one of us (L.
Yelles) using various MHD simulation snapshots has shown that observations at
5 wavelength points in the line plus one at the continuum should be adequate
to obtain the magnetic field vector reliably.

The computations carried out here suggest that this is also true for the
magnetic field extrapolated from vector magnetograms obtained from
filter instruments. According to Table \ref{table2} and Figs. \ref{fig2} and
\ref{fig2b} the application of $100$ m{\AA} broad filters to 5 locations in
the Fe I $6173.3$ {\AA} line and additionally to a nearby continuum position
gives an extrapolated nonlinear force-free field that differs only slightly
from the results obtained with the
full line profile. The magnetic energy is overestimated by at most
$6 \%$ for a filter width $< 100$ m{\AA} and at least 5 sampling points in
the line and one in the nearby continuum.

Noise has a somewhat larger effect on filter polarimeter measurements than on
spectropolarimetric ones, as can be judged from Table \ref{table2}. In
particular, the magnetic energy is affected, being more than $2 \times$ less
accurate than for spectropolarimetric measurements.
If we take the effect of photometric noise of
$10^{-3} I_c$ in all Stokes parameters into account we overestimate the
magnetic energy by $10 \%$. The somewhat higher magnetic energy for these
cases is probably a result of stronger currents in the photosphere, which are
created by a less accurate computation of the horizontal photospheric field
during the inversion. The total vertical magnetic field $\int |B_z| \;
dx dy$ in the photosphere is underestimated by $7 \%$ after inversion of a
set of filter images (irrespective of whether noise is applied or not). The
total vertical current $\int \left| \frac{\partial B_y}{\partial
x}-\frac{\partial B_x}{\partial y} \right| \; dx dy$ is overestimated by $42
\%$ and $262 \%$ without and with noise, respectively. Most of these
spurious currents produced by spectral line inversions
fluctuate on very small scales so that they
either are noise, or behave like that. Consequently, they are not transported
into the corona. The preprocessing routine takes care of this problem and
most of the spurious currents vanish. After preprocessing the total current
is overestimated by $2 \%$ and $9 \%$ for inversions without and with noise,
respectively. %
\subsection{Hinode-like cases}
Finally, as discussed in this and the following subsection, we add more
realism into the synthetic observations by employing parameters that are
appropriate to high resolution instruments. In this section we consider the
important case of the spectropolarimeter on Hinode (see also Section
\ref{hinode_cases}).

Taking a finite spatial resolution (pixel size 110 km) and spectral smearing
into account naturally leads to less accurate results
(rows marked Hinode/SP in Table \ref{table2}).
We find that we cannot reconstruct the magnetic
energy accurately for these cases, because small scale magnetic fields are
not adequately resolved. We get, however, a reasonable estimate of the
correct magnetic field in higher layers of the atmosphere. For the most
involved case (Hinode-like spectropolarimeter with a noise of
$10^{-3} I_c$ and a pixel size of
110 km) we find an error of $16\%, \, 5\%$ and $1\%$ at the heights 100, 400
and 800 km above $Z_{\rm ref}$, respectively. The limited resolution avoids
accurate reconstructions of low-lying small scale features, but gets the
higher lying field approximately correctly. At heights above 400 km even the
most involved and noisy nonlinear force-free reconstruction considered here
has an accuracy, which is three times better than with a potential field
reconstruction starting from a perfectly known lower boundary. %
\subsection{SO/PHI-like cases}
We consider instrumental effects appropriate to the PHI instrument on Solar
Orbiter such as the finite spectral resolution and sampling by a Fabry-Perot
interferometer, finite spatial resolution of approximately
 160 km (pixel size 80 km) on the
Sun and photon noise at a level of $10^{-3} \times \frac{1}{\sqrt{I_c}}$.
Details are given in Sect. \ref{PHI_cases}. The extrapolated magnetic field displays a
very similar spatial distribution as that based on the
inversion of ideal line profiles. On the whole, the influence of
degrading the spatial resolution to
a pixel size of $80$ km (denoted as {\bf Filter+Noise} in Figs. \ref{fig2},
\ref{fig2b}) introduces a similar level of inaccuracy in the extrapolated
field as the uncertainties in $B_x, B_y, B_z$ at the lower boundary introduced by
instrumental effects and the inversion of the line profiles. But even for
this most involved cases, with comparatively low resolution, instrument effects and noise,
the agreement with the reference is better than a potential field computed
from ideal data at the lower boundary.
The better correspondence of $\epsilon$ with the reference
value than for Hinode is likely due to the somewhat higher spatial resolution
expected for PHI. This suggests that for accurate estimates of the magnetic
energy it is more important to achieve high spatial resolution than
completely accurate line profiles. The field at 100 km above $Z_{\rm ref}$
and partly at 400 km is better
reproduced for Hinode/SP-like parameters than for the SO/PHI case.
Obviously, for this it is better to have the full line profile (in particular
when noise is included).

\section{Conclusions}
We have investigated how strongly inaccuracies in the lower magnetic
boundary, introduced by measurement and analysis errors of the Stokes
profiles influence the computation of nonlinear force-free coronal magnetic
fields.

We find that instrument effects and noise influence the horizontal
  component of the photospheric magnetic field vector stronger than the
  vertical field. We find that non-linear
  force-free fields extrapolated from
  ideal data and from the inversion of Stokes profiles deviate more strong at low
  heights. In particular, a limited spatial resolution influences the
  lowest layers the strongest.  Higher in the atmosphere we found a very good agreement
  (correlation better than $0.98$) with extrapolations from ideal data.
  We find that for an accurate estimation of the magnetic energy a high
  spatial resolution is more important than a high spectral resolution.

These basic findings apply to magnetic vector maps obtained from both,
spectropolarimetric data, such as provided by Hinode/SP, and filter
magnetographs, such as to be provided by SO/PHI or SDO/HMI.

\begin{figure}
   \centering
\includegraphics[width=8cm]{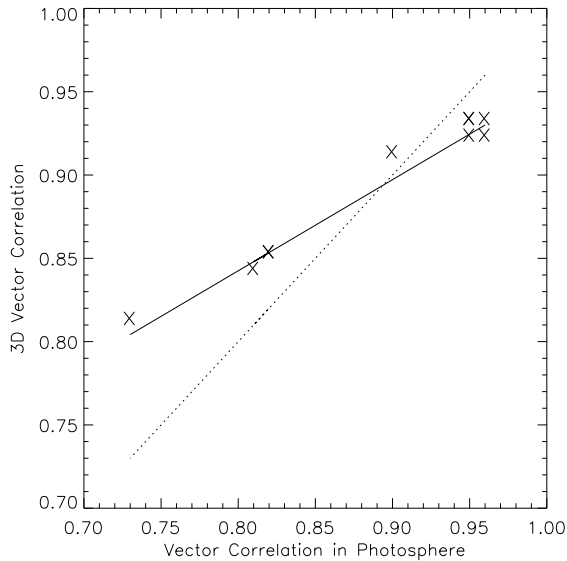}
\caption{Scatter plot displaying how errors in the photosphere affect the accuracy
of the 3D magnetic field reconstruction. Shown is  the 3D vector correlation
against the 2D vector correlation in the photosphere.
The solid line corresponds to
a linear fit and the dotted line corresponds to  identical 2D and 3D correlations.}
\label{scatter}%
\end{figure}
Finally, we would like to determine how errors in the photosphere
influence the quality of the 3D reconstruction. For this aim we
show a scatter plot in Fig. \ref{scatter} to
compare the 2D-vector correlation in the photosphere
(Table \ref{table1}) with the 3D vector correlation of the
reconstructed coronal magnetic field (Table \ref{table2}).
The solid line in Fig. \ref{scatter} shows a linear fit
to the data points and
the dotted line corresponds to equal 2D and 3D correlations.
As one can see from this figure the relation between the accuracy
of the photospheric and coronal field is linear, but not identical.
In particular, the correlation of the full 3D field drops off
more slowly than the photospheric correlation, so that extrapolations
based on a filter instrument with noise (which gives the least
accurate photospheric field) are more accurate than
the vector magnetograms they are based on.

It would be interesting to repeat this study on larger spatial scales,
reaching sizes typical of observed vector magnetograms, as soon as the
corresponding radiative MHD-simulations become
available (with a similar grid size as the simulations employed here).
There is also a need to investigate quiet Sun regions, coronal
holes and active regions separately. In particular, our results may
not apply when considering extrapolations starting from regions containing
sunspots due to their much larger Wilson depression and rather different
temperature structure, which influence the line formation height and
the line formation in general.
We have also not considered subtle, but
possibly important effects such as the evolution of the field during the scan
of an active region by the slit of a spectropolarimeter, or the evolution of
a line profile during the spectral scan of a filter instrument.

One might also consider investigating the lower layers of the solar
atmosphere, where the plasma is not force-free, in more detail and taking
non-magnetic forces into consideration for the magnetic field extrapolation. A
first step in this direction has recently been taken by
\cite{wiegelmann:etal06b}, who developed a magnetohydrostatic extrapolation
code. For an application to data we require, however, more information
regarding the nature of the non-magnetic forces (e.g., pressure gradients and
gravity).
\begin{acknowledgements}
We thank  R. Cameron, M. Sch\"ussler and A. V\"ogler for providing
us with the analysed numerical simulation snapshot. The work of T.\
Wiegelmann was supported by DLR-grant 50 OC 0501.
This work was partly supported by WCU grant No. R31-10016 from the
Korean Ministry of Education, Science and Technology.
\end{acknowledgements}

\bibliographystyle{aa}
\bibliography{tw}

\begin{thebibliography}{37}
\expandafter\ifx\csname natexlab\endcsname\relax\def\natexlab#1{#1}\fi

\bibitem[{{Aly}(1989)}]{aly89}
{Aly}, J.~J. 1989, Sol. Phys., 120, 19

\bibitem[{{Auer} {et~al.}(1977){Auer}, {House}, \& {Heasley}}]{auer:etal77}
{Auer}, L.~H., {House}, L.~L., \& {Heasley}, J.~N. 1977, Sol. Phys., 55, 47

\bibitem[{{Charbonneau}(1995)}]{charbonneau95}
{Charbonneau}, P. 1995, \apjs, 101, 309

\bibitem[{{Danilovic} {et~al.}(2008){Danilovic}, {Gandorfer}, {Lagg},
  {Sch{\"u}ssler}, {Solanki}, {V{\"o}gler}, {Katsukawa}, \&
  {Tsuneta}}]{danilovic:etal08}
{Danilovic}, S., {Gandorfer}, A., {Lagg}, A., {et~al.} 2008, A\&A, 484, L17

\bibitem[{{DeRosa} {et~al.}(2009){DeRosa}, {Schrijver}, {Barnes}, {Leka},
  {Lites}, {Aschwanden}, {Amari}, {Canou}, {McTiernan}, {R{\'e}gnier},
  {Thalmann}, {Valori}, {Wheatland}, {Wiegelmann}, {Cheung}, {Conlon},
  {Fuhrmann}, {Inhester}, \& {Tadesse}}]{derosa:etal09}
{DeRosa}, M.~L., {Schrijver}, C.~J., {Barnes}, G., {et~al.} 2009, ApJ, 696,
  1780

\bibitem[{{Frutiger} {et~al.}(2000){Frutiger}, {Solanki}, {Fligge}, \&
  {Bruls}}]{frutiger:etal00}
{Frutiger}, C., {Solanki}, S.~K., {Fligge}, M., \& {Bruls}, J.~H.~M.~J. 2000,
  A\&A, 358, 1109

\bibitem[{{Gary}(2001)}]{gary01}
{Gary}, G.~A. 2001, Sol. Phys., 203, 71

\bibitem[{{Gary} \& {Hagyard}(1990)}]{gary:etal90}
{Gary}, G.~A. \& {Hagyard}, M.~J. 1990, Sol. Phys., 126, 21

\bibitem[{{Khomenko} {et~al.}(2005{\natexlab{a}}){Khomenko}, {Mart{\'{\i}}nez
  Gonz{\'a}lez}, {Collados}, {V{\"o}gler}, {Solanki}, {Ruiz Cobo}, \&
  {Beck}}]{khomenko:etal05}
{Khomenko}, E.~V., {Mart{\'{\i}}nez Gonz{\'a}lez}, M.~J., {Collados}, M.,
  {et~al.} 2005{\natexlab{a}}, A\&A, 436, L27

\bibitem[{{Khomenko} {et~al.}(2005{\natexlab{b}}){Khomenko}, {Shelyag},
  {Solanki}, \& {V{\"o}gler}}]{khomenko:etal05a}
{Khomenko}, E.~V., {Shelyag}, S., {Solanki}, S.~K., \& {V{\"o}gler}, A.
  2005{\natexlab{b}}, A\&A, 442, 1059

\bibitem[{{Lagg} {et~al.}(2004){Lagg}, {Woch}, {Krupp}, \&
  {Solanki}}]{lagg:etal04}
{Lagg}, A., {Woch}, J., {Krupp}, N., \& {Solanki}, S.~K. 2004, A\&A, 414, 1109

\bibitem[{{Landi degl'Innocenti}(1992)}]{landi92}
{Landi degl'Innocenti}, E. 1992, {Magnetic field measurements, in: Solar
  observations: Techniques and interpretation, Ed.: Sanchez, Collados, Vazquez}
  (Cambridge University Press), 71--143

\bibitem[{{Leka} {et~al.}(2009){Leka}, {Barnes}, {Crouch}, {Metcalf}, {Gary},
  {Jing}, \& {Liu}}]{leka:etal09}
{Leka}, K.~D., {Barnes}, G., {Crouch}, A.~D., {et~al.} 2009, Sol. Phys., 139

\bibitem[{{Lites} {et~al.}(2007){Lites}, {Elmore}, {Streander}, {Hoffmann},
  {Tarbell}, {Title}, {Shine}, {Ichimoto}, {Tsuneta}, {Shimizu}, \&
  {Suematsu}}]{lites:etal07}
{Lites}, B.~W., {Elmore}, D.~F., {Streander}, K.~V., {et~al.} 2007, in New
  Solar Physics with Solar-B Mission, Vol. 369, ASP Conf. Ser., ed.
  K.~{Shibata}, S.~{Nagata}, \& T.~{Sakurai}, 55

\bibitem[{{Low} \& {Lou}(1990)}]{low:etal90}
{Low}, B.~C. \& {Lou}, Y.~Q. 1990, ApJ, 352, 343

\bibitem[{{Metcalf} {et~al.}(2008){Metcalf}, {Derosa}, {Schrijver}, {Barnes},
  {van Ballegooijen}, {Wiegelmann}, {Wheatland}, {Valori}, \&
  {McTtiernan}}]{metcalf:etal08}
{Metcalf}, T.~R., {Derosa}, M.~L., {Schrijver}, C.~J., {et~al.} 2008, \solphys,
  247, 269

\bibitem[{{Metcalf} {et~al.}(1995){Metcalf}, {Jiao}, {McClymont}, {Canfield},
  \& {Uitenbroek}}]{metcalf:etal95}
{Metcalf}, T.~R., {Jiao}, L., {McClymont}, A.~N., {Canfield}, R.~C., \&
  {Uitenbroek}, H. 1995, ApJ, 439, 474

\bibitem[{{Metcalf} {et~al.}(2006){Metcalf}, {Leka}, {Barnes}, {Lites},
  {Georgoulis}, {Pevtsov}, {Balasubramaniam}, {Gary}, {Jing}, {Li}, {Liu},
  {Wang}, {Abramenko}, {Yurchyshyn}, \& {Moon}}]{metcalf:etal06}
{Metcalf}, T.~R., {Leka}, K.~D., {Barnes}, G., {et~al.} 2006, Sol. Phys., 237,
  267

\bibitem[{Rachkowsky(1967)}]{rachkowsky:67}
Rachkowsky, D.~N. 1967, Izv. Krym. Astrofiz. Obs., 37, 56

\bibitem[{{Schrijver} {et~al.}(2008){Schrijver}, {DeRosa}, {Metcalf}, {Barnes},
  {Lites}, {Tarbell}, {McTiernan}, {Valori}, {Wiegelmann}, {Wheatland},
  {Amari}, {Aulanier}, {D{\'e}moulin}, {Fuhrmann}, {Kusano}, {R{\'e}gnier}, \&
  {Thalmann}}]{schrijver:etal08}
{Schrijver}, C.~J., {DeRosa}, M.~L., {Metcalf}, T., {et~al.} 2008, ApJ, 675,
  1637

\bibitem[{{Schrijver} {et~al.}(2006){Schrijver}, {Derosa}, {Metcalf}, {Liu},
  {McTiernan}, {R{\'e}gnier}, {Valori}, {Wheatland}, \&
  {Wiegelmann}}]{schrijver:etal06}
{Schrijver}, C.~J., {Derosa}, M.~L., {Metcalf}, T.~R., {et~al.} 2006, Sol.
  Phys., 235, 161

\bibitem[{{Shimizu}(2004)}]{shimizu04}
{Shimizu}, T. 2004, in ASP Conf. Ser. Vol. 325: The Solar-B Mission and the
  Forefront of Solar Physics, ed. T.~{Sakurai} \& T.~{Sekii}, 3

\bibitem[{{Solanki}(1987)}]{solanki1987}
{Solanki}, S.~K. 1987, PhD thesis, No.~8309, ETH, Z{\"u}rich, (1987)

\bibitem[{{Solanki} {et~al.}(2003){Solanki}, {Lagg}, {Woch}, {Krupp}, \&
  {Collados}}]{solanki:etal03}
{Solanki}, S.~K., {Lagg}, A., {Woch}, J., {Krupp}, N., \& {Collados}, M. 2003,
  Nature, 425, 692

\bibitem[{{Titov} \& {D{\'e}moulin}(1999)}]{titov:etal99}
{Titov}, V.~S. \& {D{\'e}moulin}, P. 1999, A\&A, 351, 707

\bibitem[{{Tsuneta} {et~al.}(2008){Tsuneta}, {Ichimoto}, {Katsukawa}, {Nagata},
  {Otsubo}, {Shimizu}, {Suematsu}, {Nakagiri}, {Noguchi}, {Tarbell}, {Title},
  {Shine}, {Rosenberg}, {Hoffmann}, {Jurcevich}, {Kushner}, {Levay}, {Lites},
  {Elmore}, {Matsushita}, {Kawaguchi}, {Saito}, {Mikami}, {Hill}, \&
  {Owens}}]{2008SoPh..249..167T}
{Tsuneta}, S., {Ichimoto}, K., {Katsukawa}, Y., {et~al.} 2008, Sol. Phys., 249,
  167

\bibitem[{{Unno}(1956)}]{unno:56}
{Unno}, W. 1956, Pub. Astron. Soc. Japan, 8, 108

\bibitem[{{Venkatakrishnan} \& {Gary}(1989)}]{venkatakrishnan:etal89}
{Venkatakrishnan}, P. \& {Gary}, G.~A. 1989, Sol. Phys., 120, 235

\bibitem[{{V{\"o}gler} {et~al.}(2005){V{\"o}gler}, {Shelyag}, {Sch{\"u}ssler},
  {Cattaneo}, {Emonet}, \& {Linde}}]{voegler:etal05}
{V{\"o}gler}, A., {Shelyag}, S., {Sch{\"u}ssler}, M., {et~al.} 2005, A\&A, 429,
  335

\bibitem[{{Wheatland} {et~al.}(2000){Wheatland}, {Sturrock}, \&
  {Roumeliotis}}]{wheatland:etal00}
{Wheatland}, M.~S., {Sturrock}, P.~A., \& {Roumeliotis}, G. 2000, ApJ, 540,
  1150

\bibitem[{{Wiegelmann}(2004)}]{wiegelmann04}
{Wiegelmann}, T. 2004, Sol. Phys., 219, 87

\bibitem[{{Wiegelmann} {et~al.}(2006{\natexlab{a}}){Wiegelmann}, {Inhester},
  {Kliem}, {Valori}, \& {Neukirch}}]{wiegelmann:etal06a}
{Wiegelmann}, T., {Inhester}, B., {Kliem}, B., {Valori}, G., \& {Neukirch}, T.
  2006{\natexlab{a}}, A\&A, 453, 737

\bibitem[{{Wiegelmann} {et~al.}(2005{\natexlab{a}}){Wiegelmann}, {Inhester},
  {Lagg}, \& {Solanki}}]{wiegelmann:etal05b}
{Wiegelmann}, T., {Inhester}, B., {Lagg}, A., \& {Solanki}, S.~K.
  2005{\natexlab{a}}, Sol. Phys., 228, 67

\bibitem[{{Wiegelmann} {et~al.}(2006{\natexlab{b}}){Wiegelmann}, {Inhester}, \&
  {Sakurai}}]{wiegelmann:etal06}
{Wiegelmann}, T., {Inhester}, B., \& {Sakurai}, T. 2006{\natexlab{b}}, Sol.
  Phys., 233, 215

\bibitem[{{Wiegelmann} {et~al.}(2005{\natexlab{b}}){Wiegelmann}, {Lagg},
  {Solanki}, {Inhester}, \& {Woch}}]{wiegelmann:etal05}
{Wiegelmann}, T., {Lagg}, A., {Solanki}, S.~K., {Inhester}, B., \& {Woch}, J.
  2005{\natexlab{b}}, A\&A, 433, 701

\bibitem[{{Wiegelmann} \& {Neukirch}(2003)}]{wiegelmann:etal03}
{Wiegelmann}, T. \& {Neukirch}, T. 2003, Nonlinear Proc. Geophys., 10, 313

\bibitem[{{Wiegelmann} \& {Neukirch}(2006)}]{wiegelmann:etal06b}
{Wiegelmann}, T. \& {Neukirch}, T. 2006, A\&A, 457, 1053

\end{thebibliography}

\end{document}